*submit stencil*

# The early high-energy afterglow emission from Short GRBs

HE HaoNing[†] & WANG XiangYu[†]

[†]Department of Astronomy, Nanjing University, Nanjing 210093, China

**We calculate the high energy afterglow emission from short Gamma-Ray Bursts (SGRBs) in the external shock model. There are two possible components contributing to the high energy afterglow: electron synchrotron emission and synchrotron self-Compton (SSC) emission. We find that for typical parameter values of SGRBs, the early high-energy afterglow emission in 10 MeV-10 GeV is dominated by synchrotron emission. For a burst occurring at redshift $z = 0.1$, the high-energy emission can be detectable by Fermi LAT if the blast wave has an energy $E \geq 10^{51}$ergs and the fraction of electrons energy $e_e \geq 0.1$. This provides a possible explanation for the high energy tail of SGRB 081024B.**

Gamma-Ray burst

## 1. Introduction

Recently, Fermi LAT telescope detected an increasing count rate associated with GRB 081024B, which is a short GRB with a duration of about 800 ms [1]. The emission from the point source was seen up to 3GeV, in the first 5 s after the trigger [2]. Nakar (2007) [3] has discussed the high energy afterglow from SGRBs analytically. Here we study the high energy afterglow emission from SGRBs through numerical calculations.

In the standard fireball model, the relativistic outflow expands quasi-spherically and interacts with the ambient medium, producing an external shock that propagates into the surrounding medium. The shock-accelerated electrons will emit synchrotron radiation in the magnetic fields generated by the shock. Meanwhile, synchrotron photons will be up-scattered by the electrons producing SSC emission at higher energies. We calculate the spectra of the high-energy emission, taking into account the Klein-Nishina (KN) effect for inverse Compton scattering, and explore whether the Fermi LAT can detect the early high-energy afterglow emission from SGRBs.

The paper is organized as follows: First, we describe the dynamics of the external shock and the energy distribution of electrons in section 2. Then in section 3 we describe the SSC emission considering the Klein-Nishina (KN) cross section and the modification of

Received_; accepted _

doi: _

[†]Corresponding author (email: xywang@nju.edu.cn)

Supported by the National Basic Research Program ("973" Program) of China (Grant 2009CB824800), the Foundation for the Authors of National Excellent Doctoral Dissertations of China, the Qing Lan Project and the NCET grant.

Compton parameter $Y$. In section 4, we show spectra of SGRBs afterglow produced by external shocks and explore the detectability by Fermi LAT, and then discuss the possible origin of the high-energy tail of GRB 081024B.

## 2. Dynamics and electron energy distribution

We study spherical GRB ejecta with a total energy of $E$ and an initial Lorentz factor $g_0$ expanding into the surrounding medium with density $n$ (cgs units are taken throughout this paper). We take the total energy $E$ in the range of $10^{50} \sim 10^{51}$ ergs, and take environment of short GRBs as an constant density medium with n=$10^{-2} \sim 1$cm$^{-3}$ [3].

Solving differential dynamic equations, two dynamic phases can be found at early times: (i) Coasting phase. The shell does not decelerate significantly until it arrives at the deceleration radius, $R_{\text{dec}} = 5.4 \times 10^{16} \text{cm} E_{50}^{1/3} g_{0,2}^{-2/3} n_{-2}^{-1/3}$, where $E \equiv 10^{50} E_{50} \text{erg}$, $g_0 = 100 g_{0,2}$ and $n = 0.01 n_{-2}$. The corresponding deceleration time is $t_{\text{dec}} \approx (1+z) R_{\text{dec}} / (2 g_0^2 c) \approx (1+z) 90 s E_{50}^{1/3} g_{0,2}^{-8/3} n_{-2}^{-1/3}$. (ii) Blandford-Mackee relativistic self-similar phase [4] ($t > t_{\text{dec}}$ and $g \gg 1$). Energy conservation dictates $g(t) = 32 E_{50}^{1/8} n_{-2}^{-1/8} t_3^{-3/8} (1+z)^{3/8}$ and $R(t) = 1.3 \times 10^{17} \text{cm} E_{50}^{1/4} n_{-2}^{-1/4} t_3^{1/4} (1+z)^{-1/4}$.

As usual, we assume that magnetic field and electrons have a fraction $e_B$ and $e_e$ of the internal energy. And the initial electron distribution is assumed to be a power law as $dN'_e / dg_e \propto g_e^{-p}$. We take $e_e = 0.1 \sim 0.5$, $e_B = 10^{-2} \sim 0.1$ and $p = 2.2$.

The minimum and cooling Lorentz factors (in the ejecta frame) of electrons are respectively $g_{e,m} = 3.1 \times 10^3 f_p e_{e,-1} g_2$ and $g_{e,c} = 2.6 \times 10^5 f_Y (1+z) e_{B,-2}^{-1} g_2^{-1} t_3^{-1} n_{-2}^{-1}$, where $f_p = 6(p-2)/(p-1)$, $f_Y = 1/(1+Y)$ and $Y$ is defined as the ratio of inverse Compton to synchrotron luminosity, i.e., $Y = L_{\text{SSC}} / L_{\text{syn}}$, which will be discussed below in detail.

## 3. Synchrotron self inverse-Compton (SSC) radiation

The SSC emissivity on the basis of Thomson cross section can be obtained from the electron distribution $dN'_e / dg_e$ and the synchrotron flux density $f'_{n'_s}$ (Hereafter superscript primes represent that quantities are measured in the comoving frame of the shell) through following expressions,

$$e_{n,\text{Tho}}^{\text{SSC}} = 3 s_T \int_{g_{e,\text{min}}}^{g_{e,\text{max}}} dg_e \frac{dN'_e}{dg_e} \int_0^1 dx g(x) f'_{n'_s} \quad (1),$$

Where $g_{e,\text{min}} = \min[g_{e,c}, g_{e,m}]$ and $g(x) = 1 + x - 2x + 2x \ln x$ [5].

KN cross section is more accurate for high energy emission. The KN limit frequency for photons scattered by an electron with a Lorentz factor $g_e$ is $h n'_{\text{KN}} = m_e c^2 / g_e$. Since at early times most of the SSC energy is emitted by $\sim g_{e,c}$ electrons which up-scatter seed photons with frequency $\sim n'_c = 3.0 \times 10^5 \text{eV} f_Y^2 (1+z)^2 e_{B,-2}^{-3/2} g_2^{-5} n_{-2}^{-3/2} t_2^{-2}$ (i.e. electrons are in the slow cooling regime). The KN limit frequency is $n'_{\text{KN}}(g_{e,c}) = 2.0 \text{eV} f_Y^{-1} (1+z)^{-1} e_{B,-2} t_2 g_2^3 n_{-2} \ll n'_c$, which shows that KN limit cannot be neglected. Therefore, it is necessary to consider SSC emissivity based on a full KN cross section as following,

$$e_{n,\text{KN}}^{\text{SSC}} = 3 s_T \int_{g_{e,\text{min}}}^{g_{e,\text{max}}} dg_e \frac{dN'_e}{dg_e} \int_{n'_{s,\text{min}}}^{\infty} dn'_s \frac{n'_s f'_{n'_s}}{4 g n_s^2} g(x,y) \quad (2),$$

where $g_{e,\text{min}} = \max[\min[g_{e,c}, g_{e,m}], h n'/(m_e c^2)]$, $n'_{s,\text{min}} = n' m_e c^2 / 4[g_e (g_e m_e c^2 - h\nu')]$, $x = 4 g_e h n'_s / m_e c^2$, $y = h n'/[x(g_e m_e c^2 - h n')]$ and

$$g(x,y) = 2y \ln y + (1+2y)(1-y) + \frac{1}{2} \frac{x^2 y^2}{(1+xy)} (1-y).$$

Here we use the energy distribution of elec-trons and synchrotron flux density as described in Huang et al. (2000) [6], considering the correction of the cooling Lorentz factor by Compton parameter $Y$, but neglecting its possible influence on the electron distribution index and the synchrotron spectrum due to the SSC scattering in the Klein-Nishina regime since there are no electrons with $g_e > g_{e,c}$ and $Y(g_e) > 1$ in the case we consider. [8,9]

Following Sari & Esin (2001) [5], the Compton parameter considering KN cross section is described as [3]

$$Y \equiv \frac{L_{SSC}}{L_{syn}} \approx \frac{U'_{syn}(n' < n'_{KN}[\max(g_{e,c}, g_{e,m})])}{U'_B}$$

$$\approx \frac{h_{rad}h_{KN}U'_e/(1+Y)}{U'_B} = \frac{h_{rad}h_{KN}e_e}{(1+Y)e_B}$$

(3)

Where $U'_{syn}$, $U'_B$ and $U'_e$ are the comoving energy density of synchrotron radiation, magnetic field and electrons. For slow cooling, the fraction electron's energy radiated is $h_{rad} \approx (n'_m/n'_c)^{(p-2)/2}$ and the fraction of synchrotron photons below KN limit frequency is approximately

$$h_{KN} = \begin{cases} 0 & n'_{KN}(g_{e,c}) \leq n'_m \\ \left(\frac{n'_{KN}(g_{e,c})}{n'_c}\right)^{(3-p)/2} & n'_m < n'_{KN}(g_{e,c}) < n'_c \\ 1 & n'_{KN}(g_{e,c}) \geq n'_c \end{cases} \quad (4)$$

For fast cooling $h_{rad} = 1$ and

$$h_{KN} = \begin{cases} 0 & n'_{KN}(g_{e,m}) \leq n'_c \\ \left(\frac{n'_{KN}(g_{e,m})}{n'_m}\right)^{1/2} & n'_c < n'_{KN}(g_{e,m}) < n'_m \\ 1 & n'_{KN}(g_{e,m}) \geq n'_m \end{cases} \quad (5)$$

## 4. Results

Fig 1 shows synchrotron and SSC spectra with typical parameters ($E = 10^{50}\text{erg}$, $e_e = 0.1$, $e_B = 0.01$, $p = 2.2$, $z = 0.25$ and $\Gamma_0 = 300$, $n = 0.01$) for SGRBs at the observer frame time 100s. Dotted lines represent spectra with

Thomson approximation, while solid lines represent spectra considering KN cross section, from which we can see that the suppression of the SSC component due to KN cross section is significant. In additional, we can see that the emission are dominated by synchrotron radiation in Fermi LAT sensitive range (20MeV~300GeV).

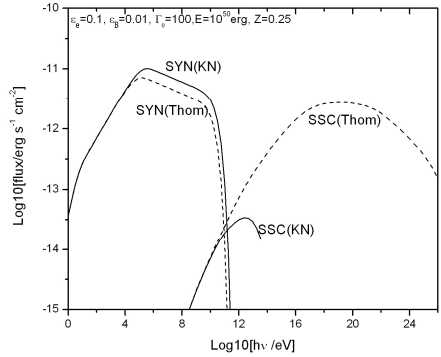

Fig. 1. Synchrotron and SSC spectra (using Thomason cross section (dashed lines) and Klein-Nishina expression (solid lines)) from external shock for a burst with parameters $E = 10^{50}\text{erg}$, $e_e = 0.1$, $e_B = 0.01$, $p = 2.2$, $z = 0.25$ and $\Gamma_0 = 100$, $n = 0.01$ at the time of 100s.

We now check whether the afterglow emission can be detected by Fermi LAT with parameters in normal range $e_e \leq 0.5$, $e_B \leq 0.1$, $E \leq 10^{51}\text{erg}$, $n = 0.01 \sim 1$, $p = 2.2$, $\Gamma_0 = 100$ for a burst occurring at $z = 0.25$. The results are shown in fig. 2, from which we can see that the high energy afterglow emission from external shock for SGRBs with normal parameters can not be detected by Fermi LAT if it is not near enough, which is consistent with the conclusion in Nakar 2007[3].

We then choose a relatively nearby burst at $z = 0.1$ and take $e_e = 0.5$, $e_B = 0.01$, $E = 10^{51}\text{erg}$, $n = 0.01$, 1, $p = 2.2$ and $\Gamma_0 = 100$, 300. Fig. 3 shows that afterglows form external shocks can be detected by Fermi LAT at different time intervals. Solid and dashed lines represent the fluence of SGRB external shock afterglow due to an outflow with a larger initial Lorentz factor $\Gamma_0 = 300$ which results in a shorter deceleration time. We can see that the afterglow can be detected by Fermi LAT, with the fluence about $7.0 \times 10^{-7} \text{ergcm}^{-2}$ for $n = 0.01$ and $7.5 \times 10^{-7} \text{ergcm}^{-2}$ for $n = 1$.

In Fig. 4., we model the extended high-energy emission from GRB081024B detected by Fermi LAT [2] between 1 and 3 s after the burst with the afterglow synchrotron emission model. We find that the afterglow synchrotron emission with the following set of parameters ($e_e = 0.3$, $e_B = 0.0001$, $n = 0.004$, $z = 0.1$, $p = 2.2$, $E = 10^{51} \text{erg}$ and $\Gamma_0 = 650$) can fit the flux level of the extended high-energy emission. This is different from the afterglow synchrotron self-Compton emission model as suggested by Corsi et al. [10]. In Fig. 5, we show that the afterglow synchrotron X-ray emission with the above parameters does not violate the non-detection by Swift/XRT between 70.3 ks and $1.3 \times 10^6$ s after the burst [11].

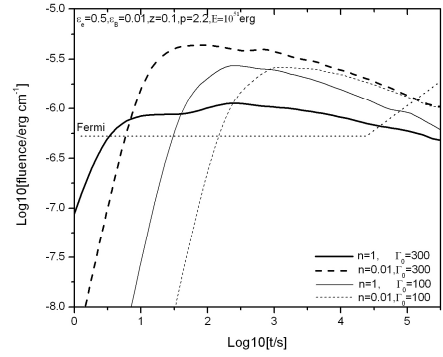

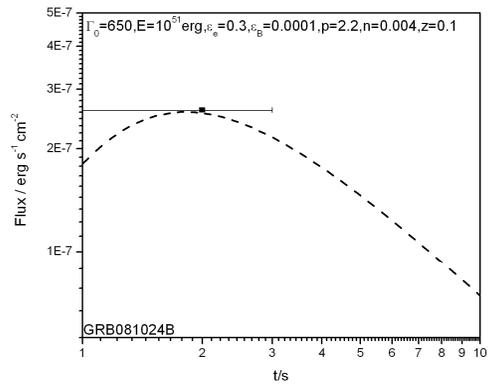

Fig. 3. The time integral fluence the SGRB afterglow emission in [20MeV, 300GeV] are presented in the solid, dashed, dash-dotted and dash-dot-dotted lines for different $\Gamma_0$ and $n$ with $z = 0.1$, $p = 2.2$, $e_e = 0.5$, $e_B = 0.01$ and $E = 10^{51} \text{erg}$. The dotted line is the sensitivity of Fermi LAT.

Fig. 4. Modelling the extended high-energy emission detected by Fermi LAT beyond the prompt keV-MeV emission with the afterglow synchrotron emission (the dashed line) for parameters $e_e = 0.3$, $e_B = 0.0001$, $n = 0.004$, $z = 0.1$, $p = 2.2$, $E = 10^{51} \text{erg}$ and $\Gamma_0 = 650$. The solid horizontal line represents the flux level at 1 GeV between 1 s and 3 s.

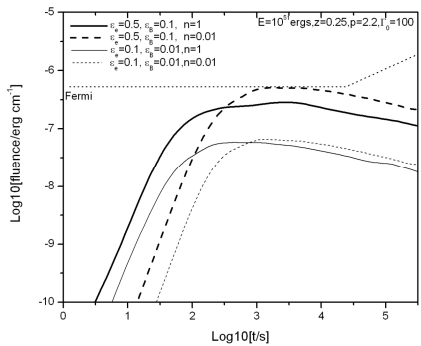

Fig. 2. The time integral fluence of the SGRB afterglow emission in [20MeV, 300GeV] are presented in the solid, dashed lines for different for different $e_e$, $e_B$ and $n$ with $z = 0.25$, $p = 2.2$, $E = 10^{51} \text{erg}$ and $\Gamma_0 = 100$. The dotted line is the sensitivity of Fermi LAT[7].

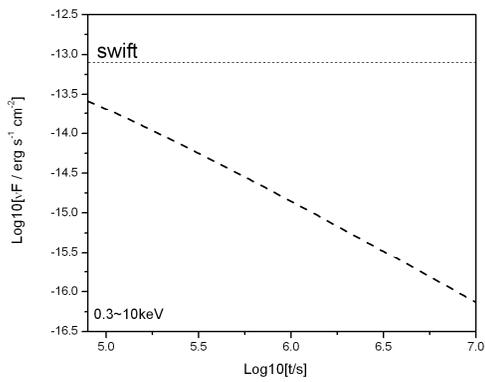

Fig. 5. The afterglow synchrotron emission (the dashed line) in the 0.3-10keV XRT band with parameters taken in Fig 4. The dotted line represents the detection sensitivity of swift/XRT for GRB081024B at time between 70.3 ks and $1.3 \times 10^6$ s after the burst, which is about $8 \times 10^{-14}$ erg cm$^{-2}$ s$^{-1}$[10, 11].

## 5. Conclusion and discussion

Since Thomson cross section is not accurate to describe the high energy emission for SGRBs, one should take KN cross section, which greatly suppresses the high energy emission from the SSC component. The afterglow emission from external shock of SGRBs in the range of Fermi LAT is dominated by synchrotron radiation. Generally, the afterglow of SGRBs located at $z \geq 0.25$ can not be detected by Fermi LAT for typical parameter values with burst energy below $10^{51}$ erg, but may be detected for more nearby bursts. SGRB 081024B might be at such distance and its high-energy tail might be due to the early afterglow synchrotron emission.